# Terahertz optoelectronics of quantum rings and nanohelices

T.P. Collier[1,*], A.M. Alexeev[1], C.A. Downing[2], O.V. Kibis[3], M.E. Portnoi[1,4, #]

[1]School of Physics, University of Exeter, Stocker Road, Exeter, EX4 4QL, United Kingdom

[2]Department of Theoretical Physics, Universidad Autonoma de Madrid, 28049 Madrid, Spain

[3]Department of Applied and Theoretical Physics, Novosibirsk State Technical University, Novosibirsk 630092, Russia

[4]ITMO University, St. Petersburg 197101, Russia

[*]e-mail: thomaspierrecollier@gmail.com

[#]e-mail: m.e.portnoi@exeter.ac.uk

**Abstract.** We outline a range of proposals on using quantum rings and nanohelices for terahertz device implementations. We show that an Aharonov-Bohm quantum ring system and a double-gated quantum ring system both permit control over the polarization properties of the associated terahertz radiation. In addition, we review the superlattice properties of a mathematically similar system, that of a nanohelix in external electric fields, which reveals negative differential conductance.

## 1. INTRODUCTION

The aptly named THz gap is a narrow region of the electromagnetic spectrum for which practical and portable technologies lack the ability to produce or detect coherent radiation [1,2]. Bridging this gap constitutes one of the trickiest problems of modern applied physics and offers potentially diverse device applications spanning from noninvasive biomedical imaging to stand-off detection of plastic explosives [3]. In this paper, we present several original proposals of using non-simply connected and chiral nanostructures as tunable active elements in optoelectronic devices operating in the sought-after THz frequency range [4-10].

## 2. QUANTUM RINGS

The progress in epitaxial growth techniques has led to the burgeoning realm of quantum ring (QR) physics [11]. The fascination with these novel nanostructures originates from the range of exotic quantum phenomena predominantly resulting from the Aharonov-Bohm effect [12,13]. Piercing a quantum ring by a magnetic flux $\phi$ yields the single-electron energy spectrum

$$\varepsilon_m(\phi) = (m + \phi/\phi_0)^2 \varepsilon_1(0), \qquad (1)$$

where $\phi_0 = h/e$ is the flux quanta, and $m \in \mathbb{Z}$ denotes the angular momentum quantum number. For a typical semiconductor ring of radius $R \sim 10$nm and electron effective mass $m_e^* = 0.07 m_0$, the energy scale $\varepsilon_1(0) = \hbar^2/2m_e^* R^2$ lies in the THz range [14,15]. The energy spectrum exhibits Aharonov-Bohm oscillations in magnetic flux with a period of $\phi_0$, and a half-integer number of flux quanta leads to a degeneracy of the electron energy levels corresponding to $m$ differing by one. This degeneracy can be easily lifted by an electric field **E** applied in the plane of the ring, which breaks the axial symmetry of the system. As such, it can be shown [4, 5] that the lowest two energy levels become separated by a gap proportional to the electric field strength $\Delta\varepsilon(\phi = \phi_0/2) = eER$. A consequence of even relatively mild electric fields $eER \sim 0.1\varepsilon_1(0)$ is the suppression of Aharonov-Bohm ground state energy oscillations. However, other quantities such as the ring's dipole moment and polarization-dependent selection rules maintain the oscillatory behavior.

Consider an electron occupying the $n^\text{th}$ state $\Psi_n$ of an infinitely thin neutral ring. The projection of the dipole moment direction on the lateral electric field is,

$$P_n = eR \int |\Psi_n|^2 \cos\varphi \, d\varphi \qquad (2)$$

where the angular coordinate $\varphi$ is measured from the horizontal axis defined by the electric field direction. In the absence of an electric field, each state is characterized by a value of $m$ ($\Psi_m = e^{im\varphi}/\sqrt{2\pi}$) and the corresponding charge density is uniformly spread over the ring. In the presence of a weak field $eER \ll \varepsilon_1(0)$ and zero magnetic field, the $m = 0$ ground state picks up only tiny contributions from wave functions with $m \neq 0$ and the charge distribution remains predominantly uniformly spread. However, at a degeneracy-inducing value of magnetic flux $\phi = \phi_0/2$, the ground state $\varphi$-dependence is well described by $\sin(\varphi/2)$, resulting in a shift of the charge density distribution *against* the electric field. Thus, upon changing flux, the ground state goes from unpolarized to strongly polarized with a dipole moment oscillation period of $\phi_0$.

Introducing finite temperature into the model requires considering the non-zero occupations of excited states. Calculations after thermal averaging over all states suggest that magneto-oscillations of the dipole moment for nanoscale rings are only observable at temperatures below 2K. The system's inter-level optical transitions, however, are less sensitive to the partial occupation of excited states at finite temperature. The transition dipole matrix element $P_{if}(\theta)$ for linearly polarized radiation incident onto the QR plane dictates the transition rate between initial ($i$) and final ($f$) states $T_{if}(\theta) \propto |P_{if}(\theta)|^2$, and for the considered ring takes the form,

$$P_{if}(\theta) = eR \int \Psi_f^* \Psi_i \cos(\theta - \varphi)\, d\varphi \qquad (3)$$

where $\theta$ is the angle between the projection of linearly polarized radiation onto the plane of the QR and the direction of the lateral electric field. Away from the degeneracies, transitions between ground and first excited state have no linear polarization. At $\phi = \phi_0/2$, the transition rate with $\theta = \pi/2$ polarization ($T_{01}(\pi/2) \equiv T_\perp$) reaches its maximum possible value, whereas transitions with $\theta = 0$ are forbidden ($T_{01}(0) \equiv T_\parallel = 0$). Such oscillations of the degree of polarization and strong optical anisotropy do not depend on temperature.

Magneto-oscillations of both the dipole moment of the ring and the degree of polarization of inter-level transitions are plotted in Fig. 1.

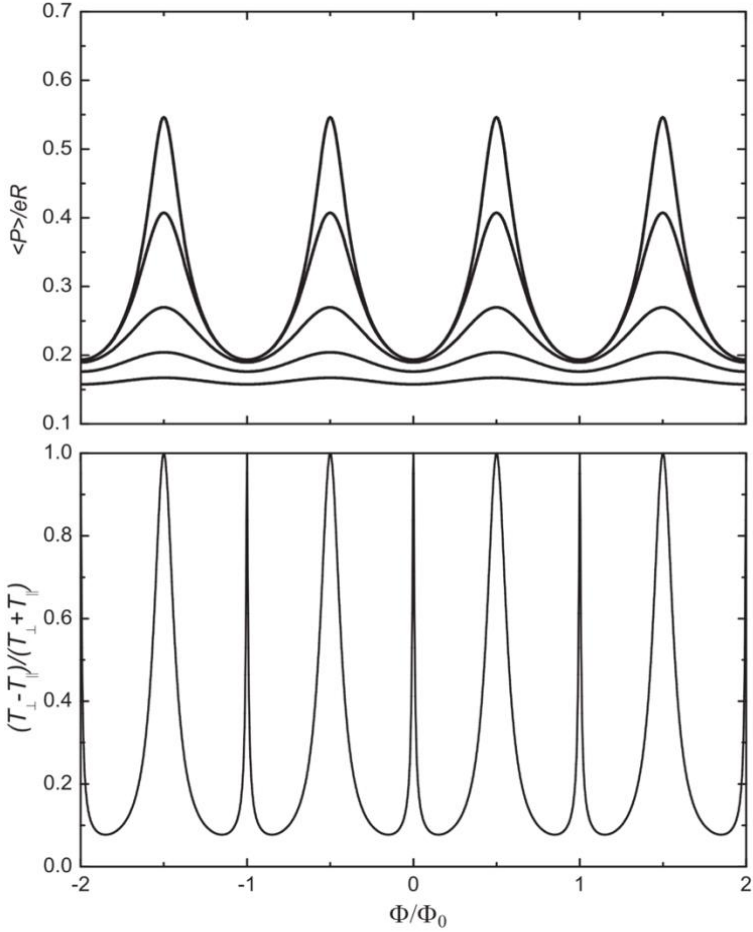

Fig. 1. The upper panel plots magneto-oscillations of the dipole moment of a ring for temperatures ranging from $0.01\varepsilon_1(0)/k_B$ (upper curve) to $0.41\varepsilon_1(0)/k_B$ (lower curve) with increments of $0.1\varepsilon_1(0)/k_B$. The lower panel shows magneto-oscillations of the degree of polarization for transitions between the two lowest energy levels. The subscripts ⊥ and ∥ denote transitions polarized perpendicular and parallel to the electric field respectively. Adapted from Ref. [4].

An additional degree of control can be achieved by placing a ring in a THz cavity [6]. If the flux piercing the ring is $\phi = \phi_0/2$, we require only a small change in the in-plane electric field to tune the energy levels of the ring into resonance with a single-mode microcavity. Additionally, the QR-microcavity coupling constant $\mathcal{G}$ (for a linearly polarized cavity mode) is proportional to $P_{if}(\theta)$. Hence, in contrast to quantum dots, for such a QR system both $\Delta\varepsilon$ and $\mathcal{G}$ can be easily controlled with external electric and magnetic fields. Thus, we have shown that Aharonov-Bohm rings can act as room-temperature polarization-sensitive THz detectors and optical magnetometers, or as a system for population inversion via optical excitation of an electron into the first excited state across the semiconductor with tunable radiative transition to the ground state. If embedded in a microcavity the system has a highly controllable emission spectrum and can be used as a tunable optical modulator or as a tool for magneto-spectroscopy.

Exploiting Aharonov-Bohm effect related properties requires trapping at least one-half of the magnetic flux quantum in the ring annulus, which requires fields of several Tesla for a ~10nm radius semiconductor ring. We show that the use of ultra-strong magnetic fields can be avoided if a ring is placed between two lateral gates [7], induce a double quantum well potential along the ring

$$V(\varphi) = \beta[(2d/R)(1-\gamma)\cos(\varphi) + \gamma\cos(2\varphi)], \tag{5}$$

where $\beta$ is the potential strength from one gate at a distance $d$ from the ring and $\gamma$ characterizes the relative potential strength of the other gate ($\gamma = 1$ describes equivalent gates). The well parameters and corresponding inter-level separations are highly sensitive to the gate voltages. Our analysis for double-gated rings shows that selection rules, caused by linearly polarized excitations for inter-level dipole transitions, Eq. (3), depend on the polarization vector angle $\theta$ with respect to the gates. In Fig. 2, we plot the probabilities of transitions between the ground and first (second) excited state for polarization angles $\theta = \pi/2$ ($\theta = 0$), the inset schematically represents the selection rules for both identical and asymmetrical gate voltages. In striking difference from the planar symmetric double well, the ring geometry permits polarization-dependent transitions between the ground and second excited states, allowing this system's use in a three-level lasing scheme.

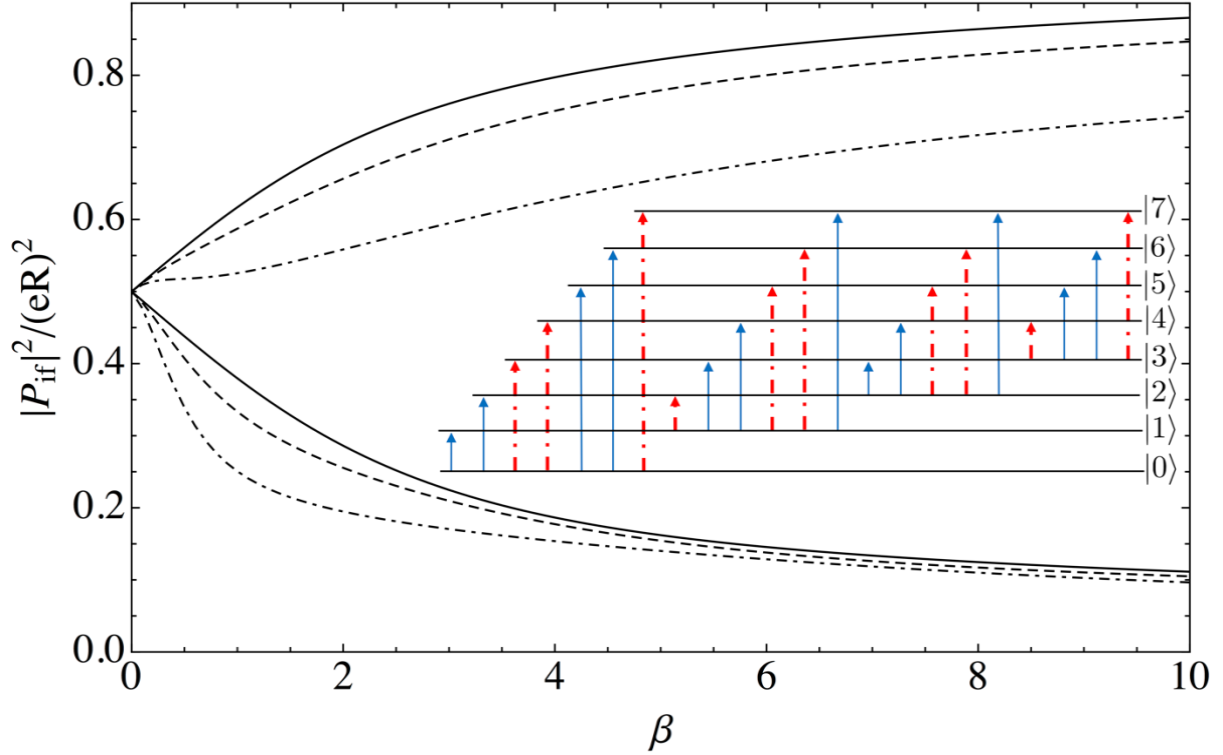

Fig. 2. Square of the dimensionless transition dipole matrix element between initial ($i$) and final ($f$) single-electron states plotted as a function of $\beta$ and with $\gamma = 1$ (full line), $\gamma = 1.003$ (dashed line), and $\gamma = 1.006$ (dot-dashed line). The upper branches denote $|P_{01}|^2/(eR)^2$ at $\theta = \pi/2$ and the lower branches denote $|P_{02}|^2/(eR)^2$ at $\theta = 0$. The inset schematically depicts the optical selection rules between the lowest eight energy eigenstates. Full blue arrows are allowed transitions when $\gamma = 1$, the red dot-dashed arrows are forbidden transitions which become allowed when $\gamma \neq 1$. Adapted from Ref. [7].

## 3. NANOHELICES

The helical form is prevalent throughout nature, not least of all contributing to the structure of DNA. Nanohelices have been fabricated in semiconductor systems [16], and for a certain type of chiral carbon nanotubes all their atoms lie on a single helix [8,9]. Here we show that a nanohelix subjected to an electric field normal to its axis behaves like a superlattice with tuneable parameters.

The problem of an ideal helix subjected to an electric field normal to its axis [8,10] is mathematically equivalent to that of a QR pierced by a flux and subjected to an in-plane electric field. The role of the flux is played by the electron momentum along the helix. The helix geometry (as with the double-gated QR system) has the advantage of not needing ultra-high magnetic fields to yield desirable physics.

Let us consider a nanohelix of radius $R$, and pitch $d$ in an external electric field $E_\perp$ normal to the helix axis. The potential energy of an electron along the helical line coordinate $s$ is then

$$V(s) = eE_\perp R \cos(2\pi s/l_0), \tag{6}$$

where $l_0 = \sqrt{4\pi^2 R^2 + d^2}$ is the length of a single coil. Clearly, $V(s)$ is periodic with period $l_0$ significantly larger than the interatomic distance, resulting in typical superlattice behaviour with tuneable electronic properties. The energy of an electron in a field-free helix is $\varepsilon_0(k) = \hbar^2 k^2/2m_e^*$. For weak electric fields $eE_\perp R \ll \varepsilon_0(2\pi/l_0)$, $V(s)$ only mixes adjacent states, which gives the spectrum

$$\varepsilon(k) = \frac{1}{2}\left[\varepsilon_0(k) + \varepsilon_0\left(|k| + \frac{2\pi}{l_0}\right)\right] \pm \frac{1}{2}\sqrt{\left[\varepsilon_0(k) + \varepsilon_0\left(|k| + \frac{2\pi}{l_0}\right)\right]^2 + eE_\perp R}. \tag{7}$$

Thus, a linearly dependent band gap $\Delta\varepsilon(k = \pm\pi/l_0) = eE_\perp R$ is opened by $E_\perp$ at the first Brillouin zone edges due to the Bragg scattering of electrons from the long-range periodic potential. In the limit of strong field $eE_\perp R \gg \varepsilon_0(2\pi/l_0)$ the electronic spectrum tends towards dispersionless flat bands well-described by energy levels in a harmonic potential [10]. When a longitudinal electric field $E_{||}$ is also applied to this system [10], such that the semiclassical motion along the $z$-axis is $k(t) = eE_{||}t/\hbar$, one can see from the drift velocity $v_d$ plotted as a function of applied field in Fig. 3 that beyond a critical field $E_{||} > E_\tau$ (where $E_\tau = \hbar/ed\tau$ and $\tau$ is a scattering time) the decreasing drift velocity implies a negative differential conductance. Accounting for tunnelling between the lowest and the next band, we find that above a certain field $E_{||} \gtrsim 4E_\tau$ the drift velocity increases once more (corresponding to N-type current-voltage characteristics).

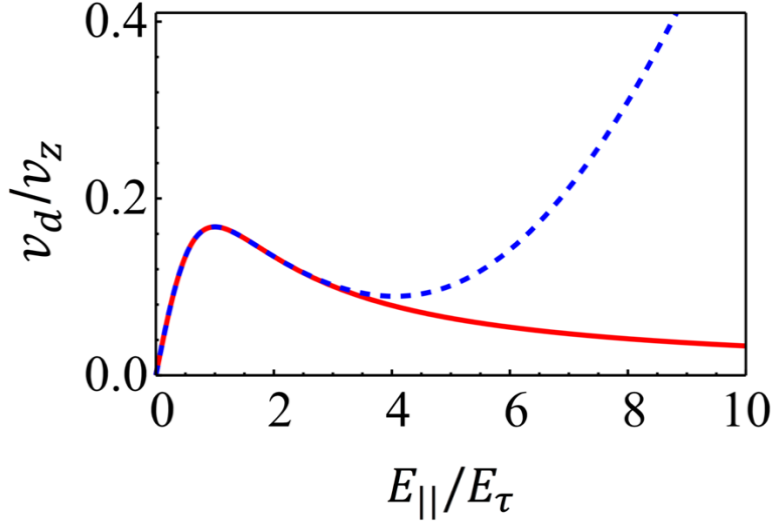

Fig. 3. Drift velocity as a function of external longitudinal electric field, without (full red line) and with (dashed blue line) the effect of tunneling from the ground band considered. We plot here for parameters $eE_\perp R = 0.4\varepsilon_0(d/2\pi)$ and $\varepsilon_0(d/2\pi)\tau = 10\hbar$, while $v_z$ is the magnitude of the velocity maxima at the Brillouin zone edge in zero external fields. Adapted from Ref. [10].

## 4. CONCLUSION

In this paper we have discussed several proposals of using ring-like and helical nanostructures in THz optoelectronics. We have shown that the high tunability of the electronic properties of these structures in external fields stems from their unusual geometries, and we have demonstrated how the corresponding phenomena may be exploited for device applications.

## ACKNOWLEDGMENTS

This work was financially supported by the EU H2020 RISE project CoExAN (TPC and MEP), the EPSRC CDT in Metamaterials XM2 (TPC), the RFBR project 18-29-19007 (OVK), and by the Government of the Russian Federation through the ITMO Fellowship and Professorship Program (MEP).